\documentclass[preprint,showpacs,preprintnumbers,amsmath,amssymb]{revtex4}

\usepackage{graphicx}

\begin{document}

\title{
Demagnifying gravitational lenses toward hunting a clue 
of exotic matter and energy 
}
\author{Takao Kitamura}
\author{Koki Nakajima}
\author{Hideki Asada} 
\affiliation{
Faculty of Science and Technology, Hirosaki University,
Hirosaki 036-8561, Japan} 

\date{\today}

\begin{abstract}
We examine a gravitational lens model inspired 
by modified gravity theories and exotic matter and energy. 
We study an asymptotically flat, static, and spherically symmetric 
spacetime that is modified in such a way that 
the spacetime metric depends on 
the inverse distance to the power of positive $n$ 
in the weak-field approximation. 
It is shown analytically and numerically that 
there is a lower limit on the source angular displacement from 
the lens object to get demagnification. 
Demagnifying gravitational lenses could appear, 
provided 
the source position
$\beta$ and the power $n$ satisfy $\beta > 2/(n+1)$ 
in the units of the Einstein ring radius 
under a large-$n$ approximation. 
Unusually, the total amplification of the lensed images, 
though they are caused by the gravitational pull, 
could be less than unity. 
Therefore, time-symmetric 
demagnification 
parts in numerical light curves 
by gravitational microlensing (F.Abe, Astrophys. J. 725, 787, 2010) 
may be evidence of an Ellis wormhole (being an example of 
traversable wormholes), but they do not always prove it. 
Such a gravitational 
demagnification 
of the light 
might be used for hunting a clue of exotic matter and energy 
that are described by an equation of state more general than 
the Ellis wormhole case. 
Numerical calculations for the $n=3$ and $10$ cases show 
maximally 
$\sim 10$ and $\sim 60$ percent depletion of the light, 
when the source position is $\beta \sim 1.1$ and 
$\beta \sim 0.7$, respectively. 
\end{abstract}

\pacs{04.40.-b, 95.30.Sf, 98.62.Sb}

\maketitle

\section{Introduction}
The bending of light 
was the first experimental confirmation of 
the theory of general relativity. 
Nowadays, gravitational lensing 
is one of the most important tools in astronomy and cosmology. 
It is widely used for investigating extrasolar planets, 
dark matter, and dark energy. 

Light bending is also of theoretical importance, in particular for studying a null structure of a spacetime. 
A rigorous form of the bending angle plays an important role 
in properly understanding a strong gravitational field  
\cite{Frittelli, VE2000, VE2002, ERT, Perlick}. 
For example, 
strong gravitational lensing in a Schwarzschild black hole 
was considered by Frittelli, Kling, and Newman \cite{Frittelli} 
and by Virbhadra and Ellis \cite{VE2000};  
Virbhadra and Ellis \cite{VE2002} later described strong gravitational lensing by naked singularities; 
Eiroa, Romero, and Torres \cite{ERT} treated 
Reissner-Nordstr\"om black hole lensing; 
Perlick \cite{Perlick} discussed lensing 
by a Barriola-Vilenkin monopole 
and also by an Ellis wormhole. 
 
One of the peculiar features of general relativity is that 
the theory admits a nontrivial topology of a spacetime;
for instance, a wormhole. 
An Ellis wormhole is a particular example of the Morris-Thorne 
traversable wormhole class \cite{Ellis, Morris1, Morris2}.
Many years ago, scattering problems in such spacetimes were discussed 
(for instance, \cite{CC, Clement}). 
One remarkable feature is that the Ellis wormhole has zero mass 
at the spatial infinity, but it causes light deflection 
\cite{CC, Clement}. 
Moreover, gravitational lensing by wormholes has been recently 
investigated as an observational probe of such an exotic spacetime 
\cite{Safonova, Shatskii, Perlick, Nandi, Abe, Toki, Tsukamoto}. 
Several forms of the deflection angle by the Ellis wormhole 
have been recently derived and often used 
\cite{Perlick, Nandi, DS, BP, Abe, Toki, Tsukamoto}. 
A reason for such differences has been clarified \cite{Nakajima, Gibbons}. 

According to recent numerical calculations by Abe \cite{Abe},  
time-symmetric demagnification parts in light curves 
could appear by gravitational microlensing effects 
of the Ellis wormhole. 
Is the time-symmetric demagnification evidence for the Ellis wormhole? 
Is is very interesting to address this question. 
One reason is that wormholes are inevitably related with 
violations of some energy conditions in physics \cite{Visser}. 
For instance, dark energy is introduced to explain 
the observed accelerated expansion of the universe 
by means of an additional energy-momentum component 
in the right-hand side of the Einstein equation. 
Furthermore, the left-hand side of the Einstein equation, 
equivalently the Einstein-Hilbert action, 
could be modified in various ways (nonlinear curvature terms, 
higher dimensions, and so on) 
inspired by string theory, loop quantum gravity, and so on. 
Because of the nonlinear nature of gravity, 
modifications to one (or both) side(s) of the Einstein equation 
might admit spacetimes significantly different from 
the standard Schwarzschild spacetime metric, 
even if the spacetime is assumed to be asymptotically flat, 
static, and spherically symmetric. 
One example is an Ellis wormhole (being an example of 
traversable wormholes). 

Inspired by a huge number of modified theories, 
this brief paper assumes, in a phenomenological sense, that 
an asymptotically flat, static, and spherically symmetric 
modified spacetime could depend on 
the inverse distance to the power of positive $n$ 
in the weak-field approximation. 
The Schwarzschild spacetime and the Ellis wormhole 
correspond to $n=1$ and $n=2$, respectively. 
Note that Birkhoff's theorem could say that cases $n \neq 1$ 
might be nonvacuum,  
if the models were interpreted in the framework of 
the standard Einstein equation. 

The slightly modified gravitational lensing 
in modified gravity theories--
such as a fourth order $f(R)$ gravity theory--
has attracted interests (e.g., Refs. \cite{Capozziello, Horvath, Mendoza}). 
It has been shown that 
the total magnification of the lensed images 
is stable and always larger than unity 
against a small spherical perturbation of 
the Schwarzschild lens \cite{Asada2011}. 
This suggests that demagnifying gravitational lenses 
would need a significantly modified structure of the spacetime. 
The main purpose of this paper is 
to discuss demagnifying gravitational lenses 
due to significantly modified spacetimes.

We take the units of $G=c=1$ throughout this paper.

\section{Modified spacetime model and 
modified deflection angle of light}
This paper assumes that 
an asymptotically flat, static, and spherically symmetric 
modified spacetime could depend on 
the inverse distance to the power of positive $n$ 
in the weak-field approximation. 
We consider light propagating through a four-dimensional spacetime, 
though the whole spacetime may be higher-dimensional. 
The four-dimensional spacetime metric is expressed as 
\begin{equation}
ds^2=-\left(1-\frac{\varepsilon_1}{r^n}\right)dt^2
+\left(1+\frac{\varepsilon_2}{r^n}\right)dr^2
+r^2(d\theta^2+\sin^2\theta d\phi^2) 
+O(\varepsilon_1^2, \varepsilon_2^2, \varepsilon_1 \varepsilon_2) ,  
\label{ds}
\end{equation}
where $r$ is the circumference radius and 
$\varepsilon_1$ and $\varepsilon_2$ are small bookkeeping 
parameters in the following iterative calculations. 
Here, $\varepsilon_1$ and $\varepsilon_2$ 
may be either positive or negative, respectively. 
A negative $\varepsilon_1$ and $\varepsilon_2$ for $n=1$ 
corresponds to a negative mass (in the linearized Schwarzschild metric). 

For investigating light propagation, it is useful below to 
make a conformal transformation with a factor 
of $(1-\varepsilon_1/r^n)^{1/2}$. 
The null structure (such as the light propagation) 
is not affected by the conformal transformation. 
At the linear order of $\varepsilon_1$ and $\varepsilon_2$, 
the spacetime metric takes a simpler form:
\begin{equation}
d\bar{s}^2=-dt^2+\left(1+\dfrac{\varepsilon}{R^n}\right)dR^2
+R^2 (d\theta^2+\sin^2\theta d\phi^2) 
+O(\varepsilon^2) , 
\label{d2}
\end{equation}
where $\varepsilon \equiv n \varepsilon_1 + \varepsilon_2$ 
and 
\begin{equation}
R^2 \equiv \dfrac{r^2}{\left(1-\dfrac{\varepsilon _1}{r^n}\right)} . 
\label{R}
\end{equation}
Note that only one parameter $\varepsilon$ enters 
the conformally transformed metric. 

For this metric, one can find the Lagrangian for 
a massless particle. 
Without loss of generality, we focus on 
the equatorial plane $\theta = \pi/2$, 
since the spacetime is spherically symmetric. 
By using the constants of motion associated with 
the timelike and rotational Killing vectors, 
the deflection angle of light 
is calculated at the linear order as 
\begin{align}
\alpha&=2 \int_{R_0}^{\infty} \frac{d\phi(R)}{dR} dR -\pi 
\nonumber\\
&=\dfrac{\varepsilon}{b^n}\int_0^{\frac{\pi}{2}} \cos^n\psi d\psi 
+O(\varepsilon^2) , 
\label{alpha}
\end{align}
where $R_0$ and $b$ denote 
the closest approach and 
the impact parameter of the light ray, respectively. 
This deflection angle recovers 
the Schwarzschild ($n=1$) and Ellis wormhole ($n=2$) cases. 
For particular cases, the above (always positive) integral factor 
becomes 
\begin{align}
\int_0^{\frac{\pi}{2}} \cos^n\psi d\psi 
&=\frac{(n-1)!!}{n!!} \frac{\pi}{2} 
\quad (\mbox{even} \: n) , 
\nonumber\\
&= \frac{(n-1)!!}{n!!} 
\quad (\mbox{odd} \: n) , 
\nonumber\\
&= \frac{\sqrt{\pi}}{2} 
\frac{\Gamma\left(\frac{n+1}{2}\right)}
{\Gamma\left(\frac{n+2}{2}\right)}
\quad (\mbox{real} \: n > 0) , 
\end{align}
Henceforth, the deflection angle is denoted simply as 
$\alpha(b) = \bar\varepsilon / b^n$ by absorbing the numerical constant  
into the $\bar\varepsilon$ parameter.

\section{Modified lens equation and its solutions}
Under the thin-lens approximation, 
it is useful to consider the lens equation as \cite{SEF} 
\begin{equation}
\beta = \frac{b}{D_L} - \frac{D_{LS}}{D_S} \alpha(b) , 
\label{lenseq}
\end{equation}
where 
$\beta$ denotes the angular position of the source and 
$D_L$, $D_S$, $D_{LS}$ are the distances from the observer 
to the lens, from the observer to the source, and from the lens to
the source, respectively. 
We wish to consider significant magnification (or demagnification), 
which could occur for a source in (or near) the Einstein ring. 
The Einstein ring is defined for $\beta=0$ \cite{SEF}. 
If $\varepsilon < 0$, 
Eq. (\ref{lenseq}) has no positive roots for $\beta = 0$ 
because of the repulsive force in the particular gravity model. 
For $\varepsilon > 0$, on the other hand, 
there is always a positive root corresponding to 
the Einstein ring. 
The negative $\varepsilon$ case is of less astronomical relevance. 
Therefore, let us consider the positive $\varepsilon$ case 
(causing the gravitational pull) 
in the following.  

In units of the Einstein ring radius, 
Eq. (\ref{lenseq}) is rewritten as 
\begin{eqnarray}
\hat\beta &=& \hat\theta - \frac{1}{\hat\theta^{n}} 
\quad (\hat\theta > 0) , 
\label{lenseqP}\\
\hat\beta &=& \hat\theta + \frac{1}{(-\hat\theta)^{n}} 
\quad (\hat\theta < 0) , 
\label{lenseqM}
\end{eqnarray}
where $\hat\beta \equiv \beta/\theta_E$ and 
$\hat\theta \equiv \theta/\theta_E$ 
for the angular position of the image $\theta \equiv b/D_L$. 

Let us consider two lines defined by $Y = 1/\hat\theta^n$ and 
$Y = \hat\theta - \beta$ in the $\hat\theta - Y$ plane. 
For $\hat\theta > 0$, therefore, we have only one intersection 
of the two lines that corresponds to one image position. 
Similarly, only one image appears for $\hat\theta < 0$.

For a general positive $n$ (e.g., $n=5$), 
it is impossible to find exact solutions for 
the modified lens equation. 
To clarify the parameter dependence, we employ 
analytic but approximate methods rather than numerical calculations. 
Furthermore, in astronomy only the significantly amplified 
images become detectable in gravitational microlensing. 
Such events occur only when a source such as a distant star 
crosses the Einstein ring. 
We thus focus on such an Einstein ring-crossing case 
as $\hat\beta < 1$ in units of the Einstein ring, 
for which Eqs. (\ref{lenseqP}) and (\ref{lenseqM}) 
are solved in the Taylor series form 
with respect to $\hat\beta$. 
We obtain 
\begin{align}
\hat\theta_{+} &= 
1 + \frac{1}{n+1}\hat\beta 
+\frac{1}{2}\frac{n}{(n+1)^2}\hat\beta^2 + O(\hat\beta^3) 
\quad (\hat\theta >0) , 
\label{thetaP}
\\
\hat\theta_{-} &= 
- 1 + \frac{1}{n+1}\hat\beta 
- \frac{1}{2}\frac{n}{(n+1)^2}\hat\beta^2 + O(\hat\beta^3) 
\quad (\hat\theta < 0) . 
\label{thetaM}
\end{align}

\section{Demagnification condition}
The amplification factor denoted as $A$ is 
$|(\beta/\theta)(d\beta/d\theta)|^{-1}$, 
namely, the inverse Jacobian of the gravitational lens mapping 
between the source and image position vectors \cite{SEF}. 
By using Eqs. (\ref{thetaP}) and (\ref{thetaM}),  
the amplification factor of each image, which is denoted by 
$A_{+}$ and $A_{-}$, respectively,  
becomes 
\begin{equation}
A_{\pm} = \frac{1}{\hat\beta (n+1)} + O(\hat\beta^0) , 
\label{Apm}
\end{equation}
where a difference between $A_{+}$ and $A_{-}$ 
appears at the next order in $\hat\beta$. 
The total amplification is thus 
\begin{align}
A_{tot} &\equiv A_{+} + A_{-} 
\nonumber\\
 &= \frac{2}{\hat\beta (n+1)} + O(\hat\beta^0). 
\label{Atot}
\end{align}

For the Schwarzschild case ($n=1$), $A_{\text{tot}} = 1/\hat\beta$. 
This is always larger than unity for $\hat\beta <1$, 
in concordance with the well-known fact. 
Demagnification of the total lensed images could occur, however, if 
\begin{equation}
\hat\beta > \frac{2}{n+1} . 
\label{demag}
\end{equation}
The larger the power $n$,  the more likely the demagnification. 
One might guess that demagnification could be caused 
for a smaller $\hat\beta$, especially $\hat\beta=0$. 
However, this is not the case. 
Equation (\ref{demag}) suggests that the total demagnification 
could occur only when $\hat\beta$ is small but larger than 
the critical value $2/(n+1)$ 
under a large-$n$ approximation. 
Note that the compatibility of the assumption $\hat\beta < 1$ 
and Eq. (\ref{demag}) implies $n > 1$. 
Namely, Eq. (\ref{demag}) becomes a better approximation as 
$n$ grows larger than unity. 

\begin{figure}
\includegraphics[width=8cm]{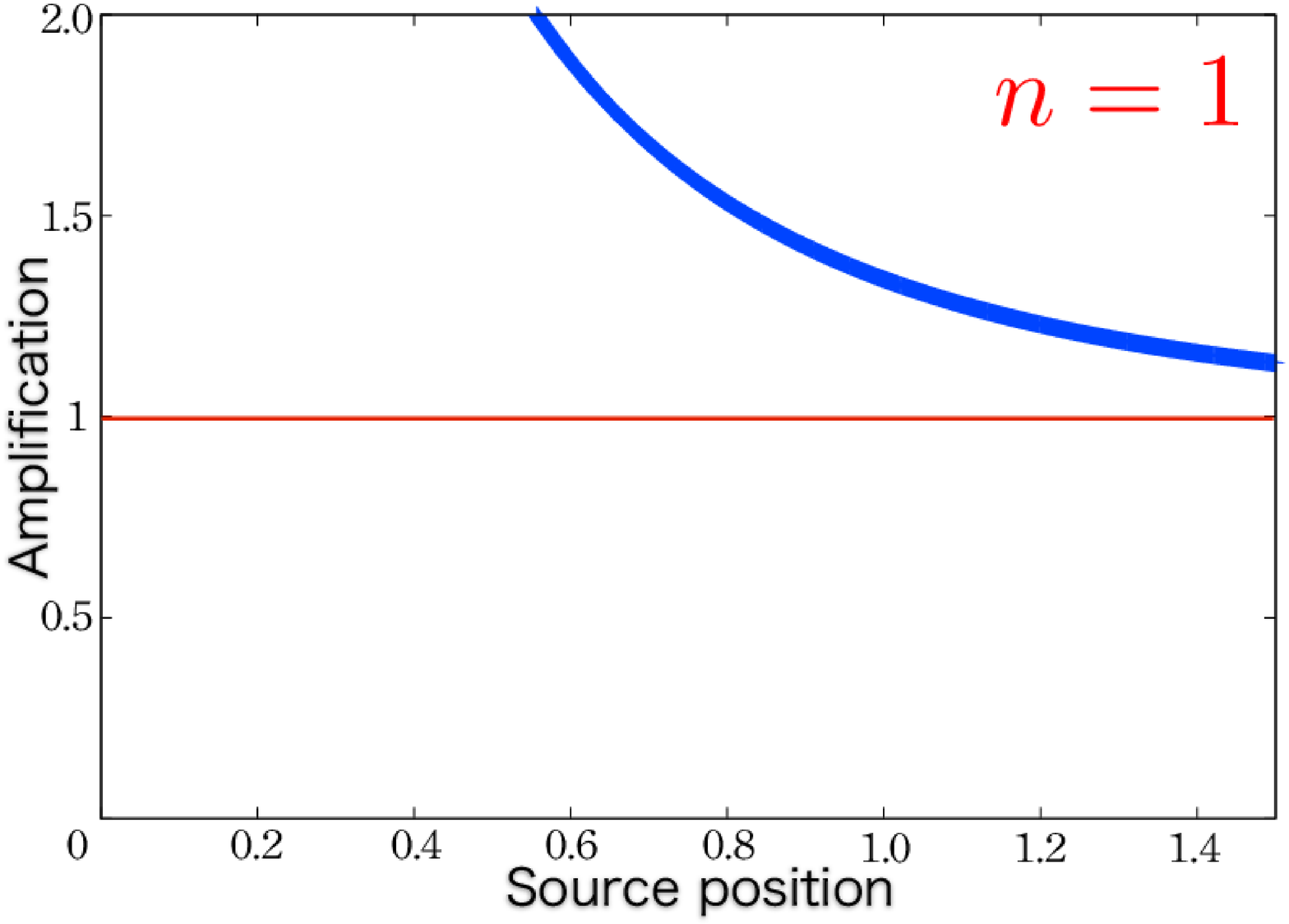}
\includegraphics[width=8cm]{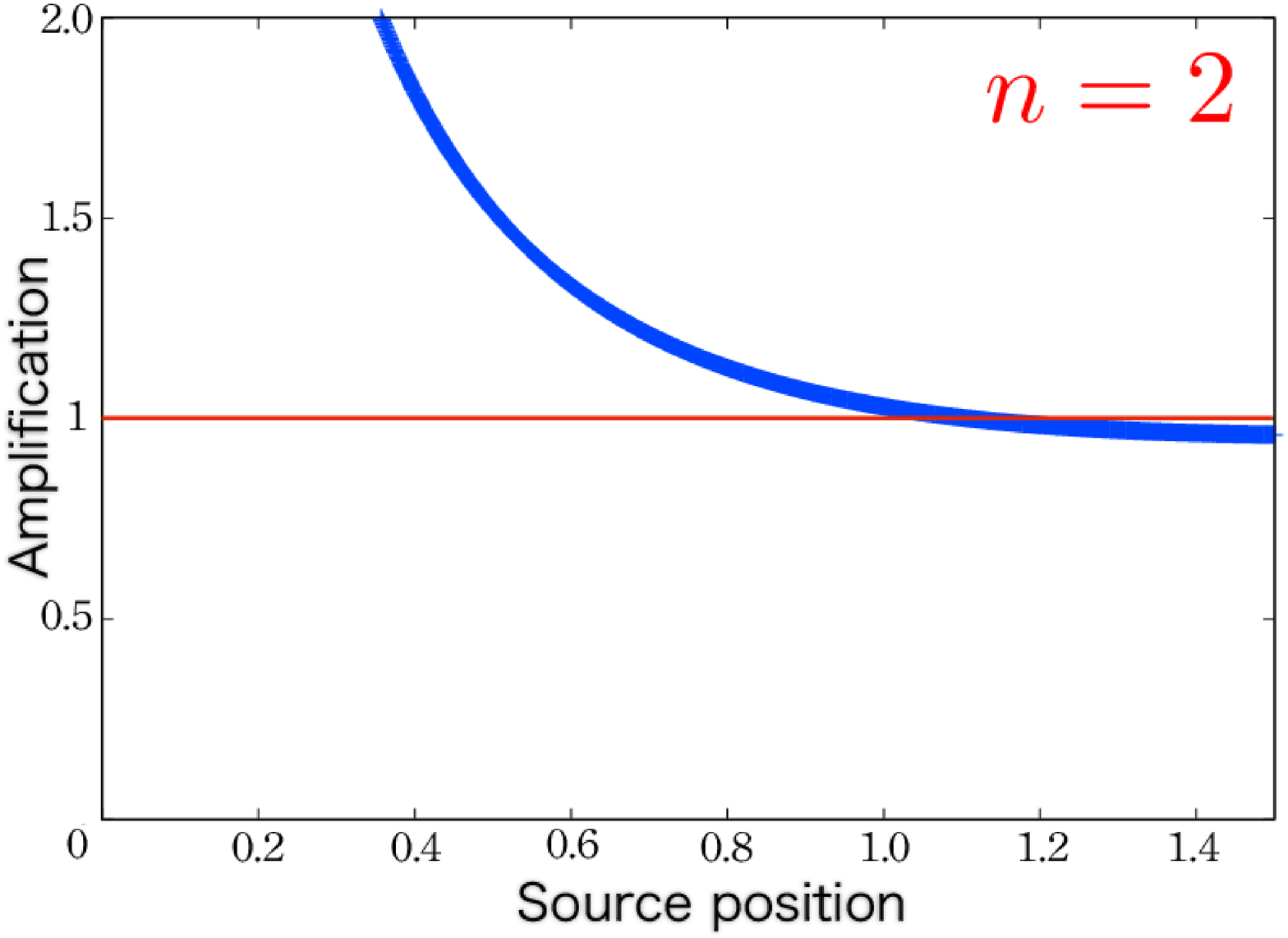}
\includegraphics[width=8cm]{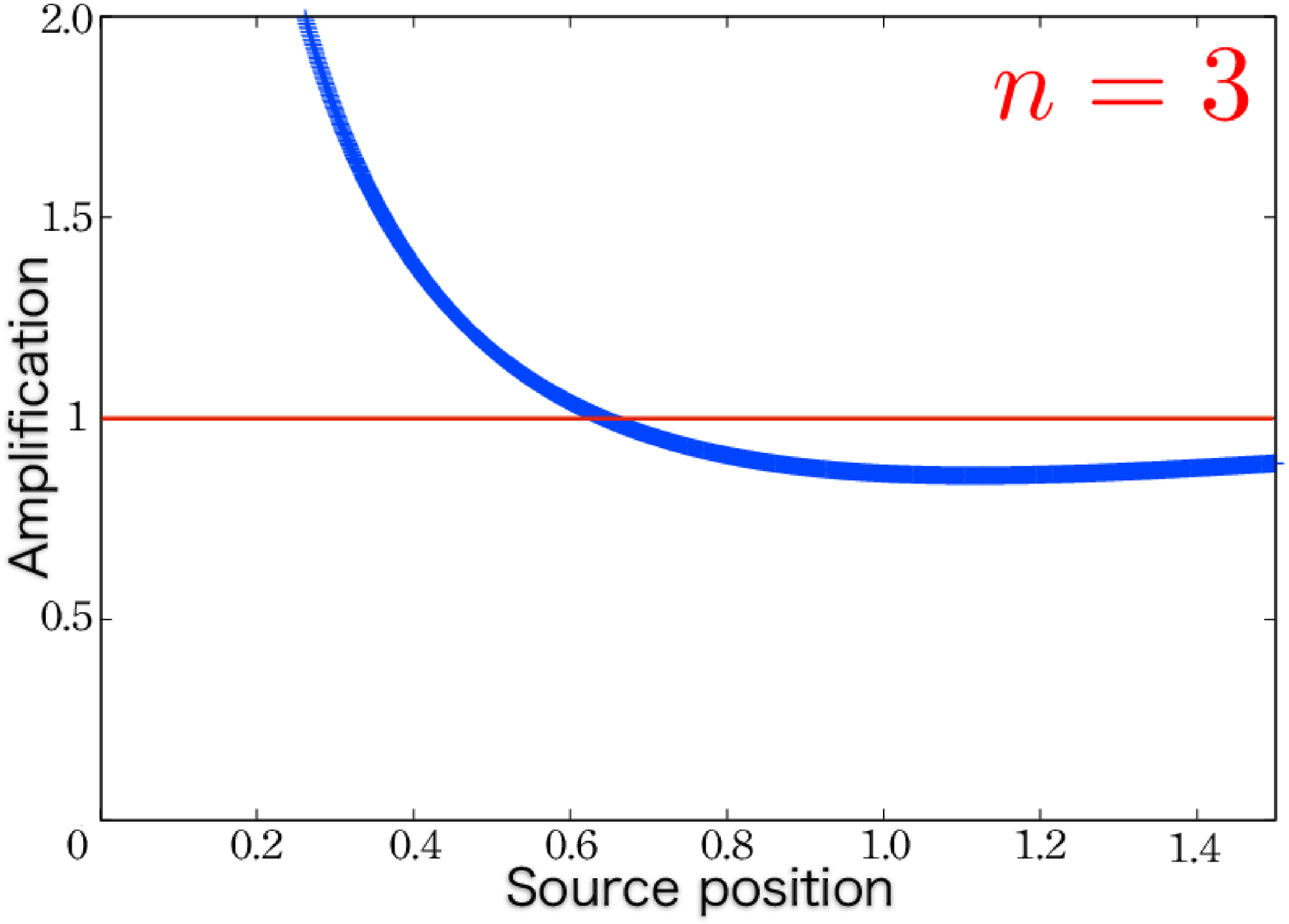}
\includegraphics[width=8cm]{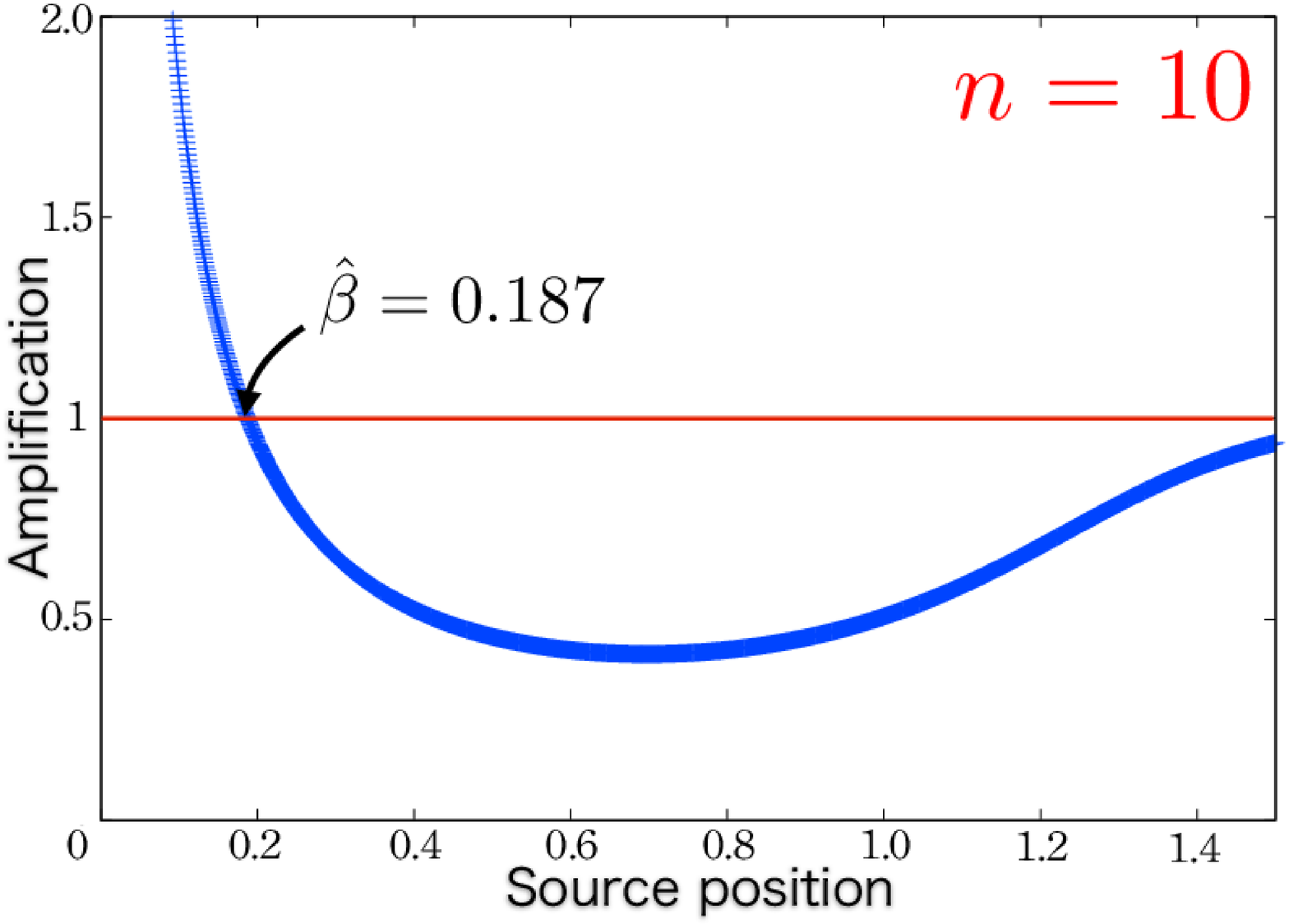}
\caption{
Total amplification factor of the lensed images 
as a function of the source position 
$\hat\beta$ for 
$n=1, 2, 3$, and $10$. 
Top left, top right, bottom left, and bottom right panels 
correspond to $n=1, 2, 3$, and $10$, respectively. 
In the case of $n=10$, 
the total amplification factor is larger than unity for 
$\hat\beta < 0.187$, whereas it is smaller for $\hat\beta > 0.187$. 
For convenience, a thin (red) line denotes $A_{\text{tot}}=1$. 
}
\label{figure-1}
\end{figure}

The above argument is based on the near-zone approximation 
($\hat\beta < 1$). 
For a test of the analytic result, 
we perform numerical calculations. 
We consider $n=10$, which might be one of the higher-dimensional models 
inspired by string theory. 
Equation (\ref{demag}) suggests that demagnification of the total lensed images 
could occur only for $\hat\beta > 2/11 = 0.182$. 
Figure \ref{figure-1} shows numerical results for 
$n=1, 2, 3$, and $10$. 
In the case of $n=10$, 
the analytic result for the critical value 
$\hat\beta = 2/11 = 0.182$ 
is in good agreement with the numerical one, $\hat\beta = 0.187$. 

\begin{figure}
\includegraphics[width=8cm]{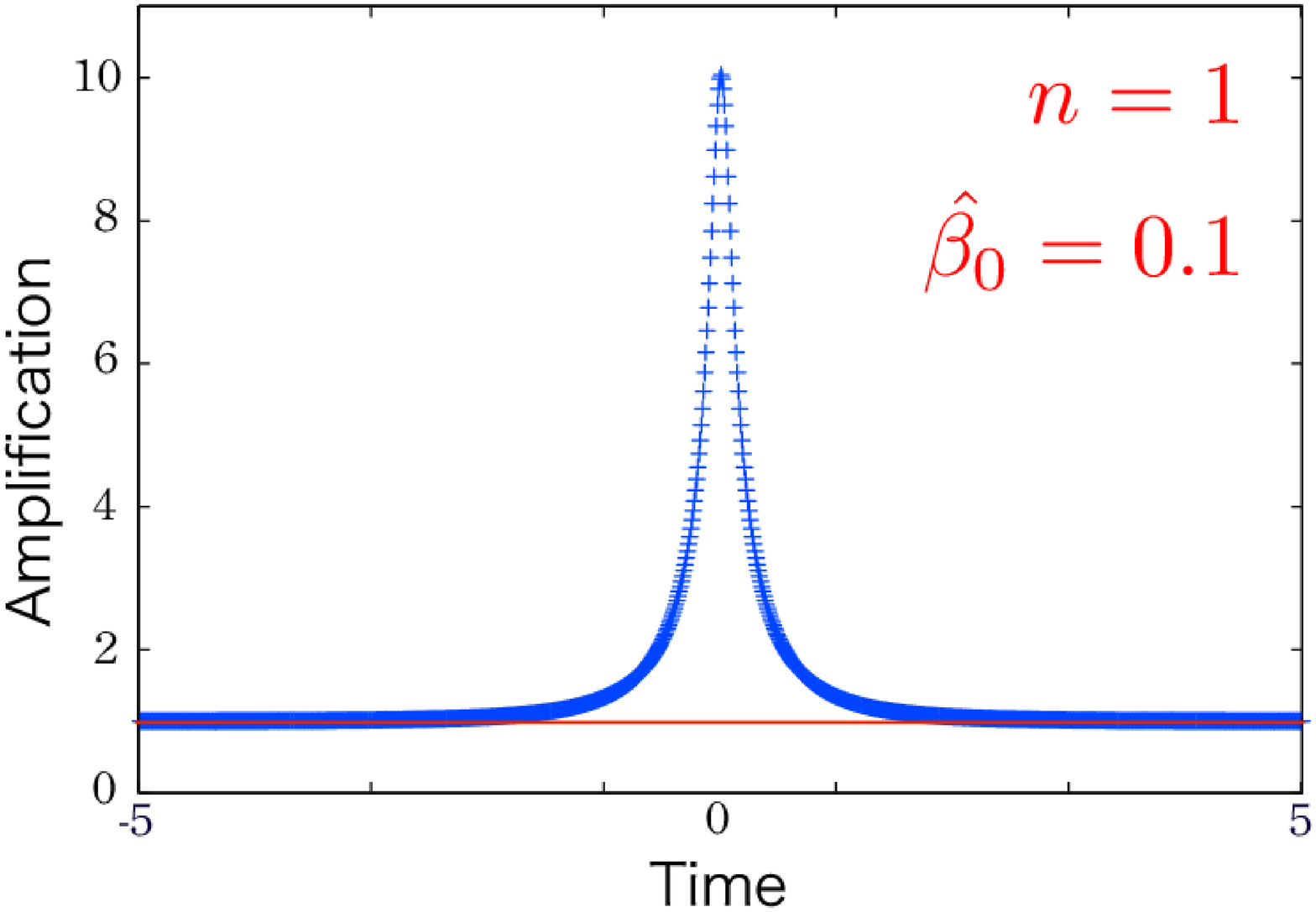}
\includegraphics[width=8cm]{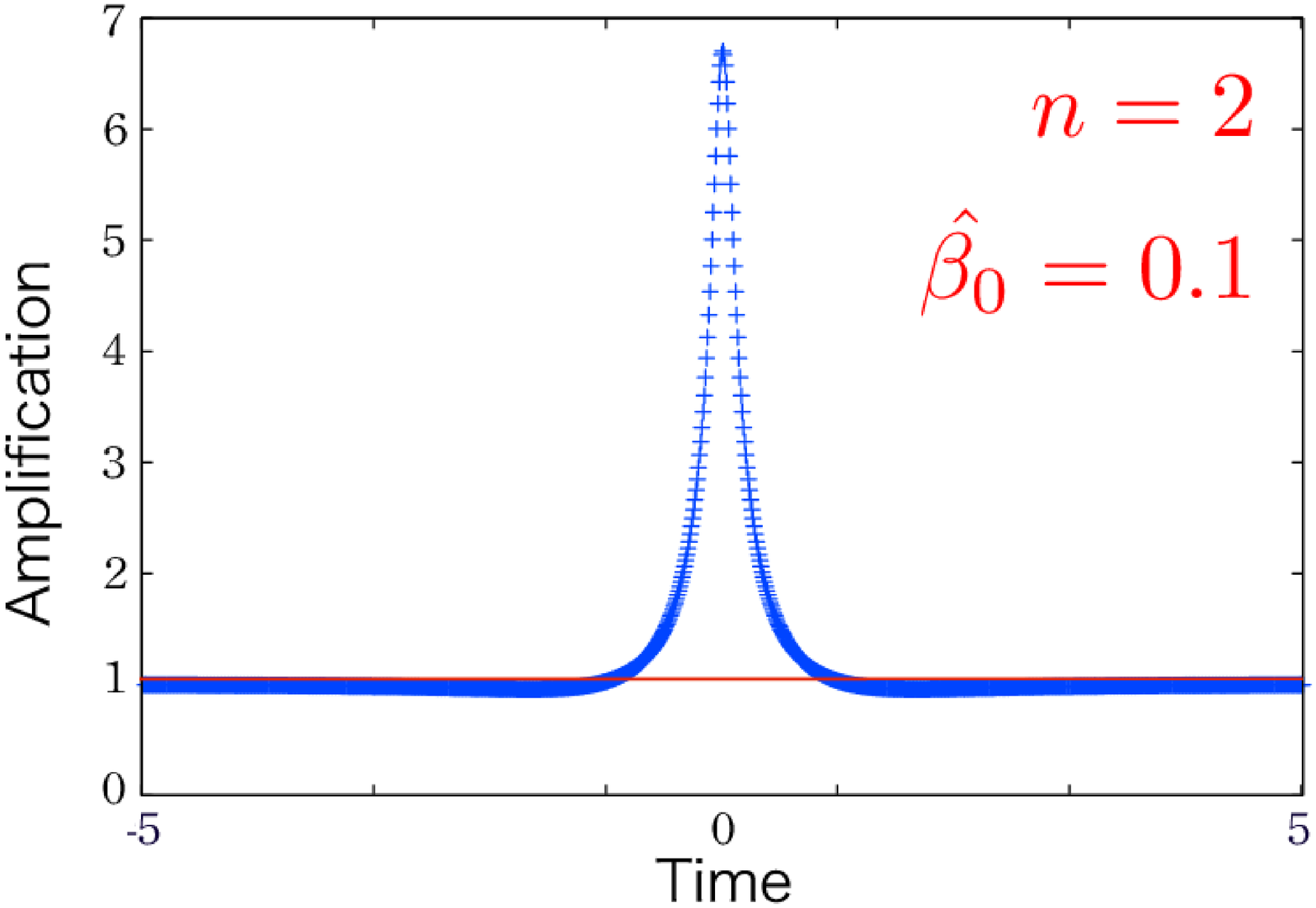}
\includegraphics[width=8cm]{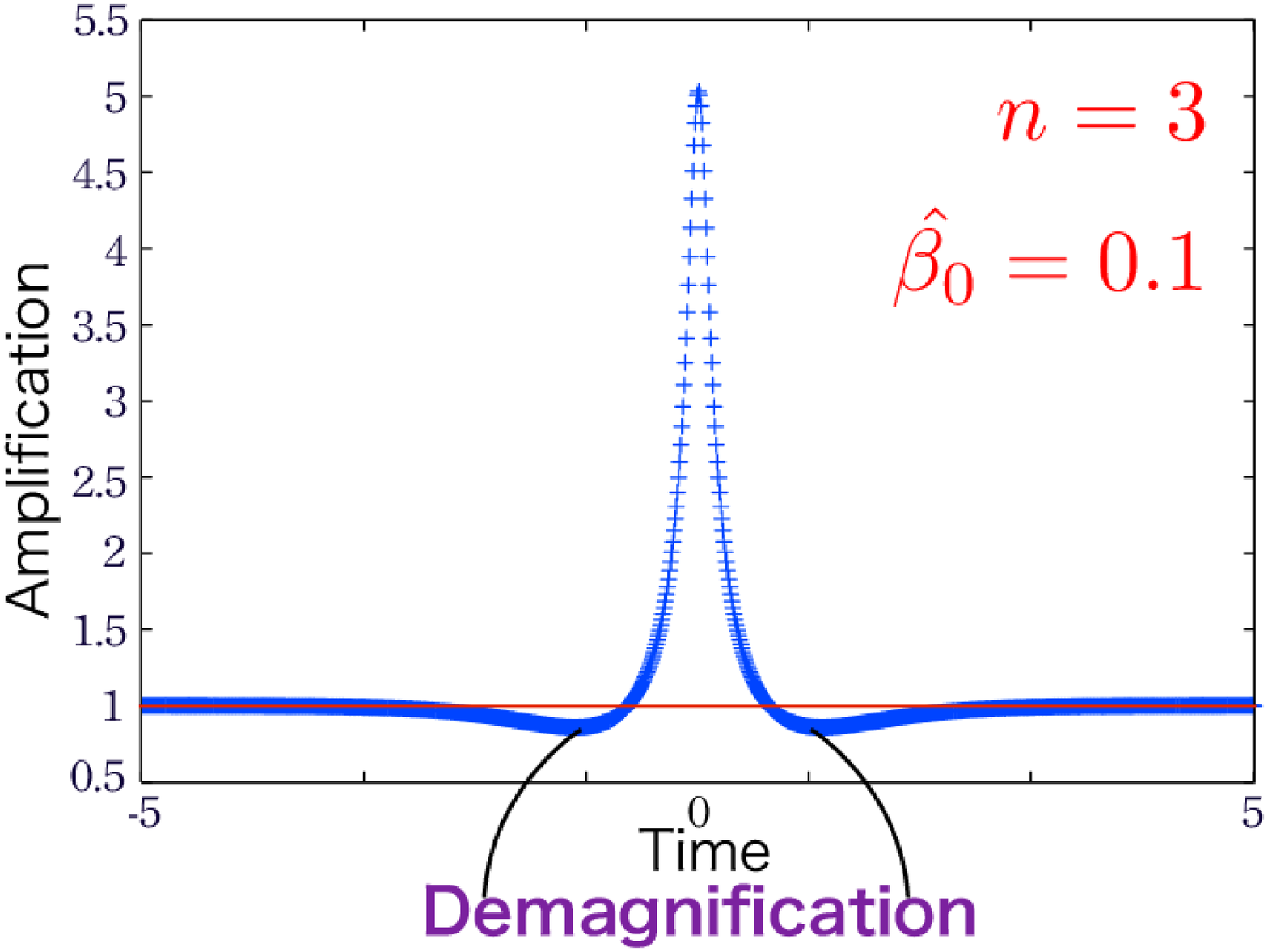}
\includegraphics[width=8cm]{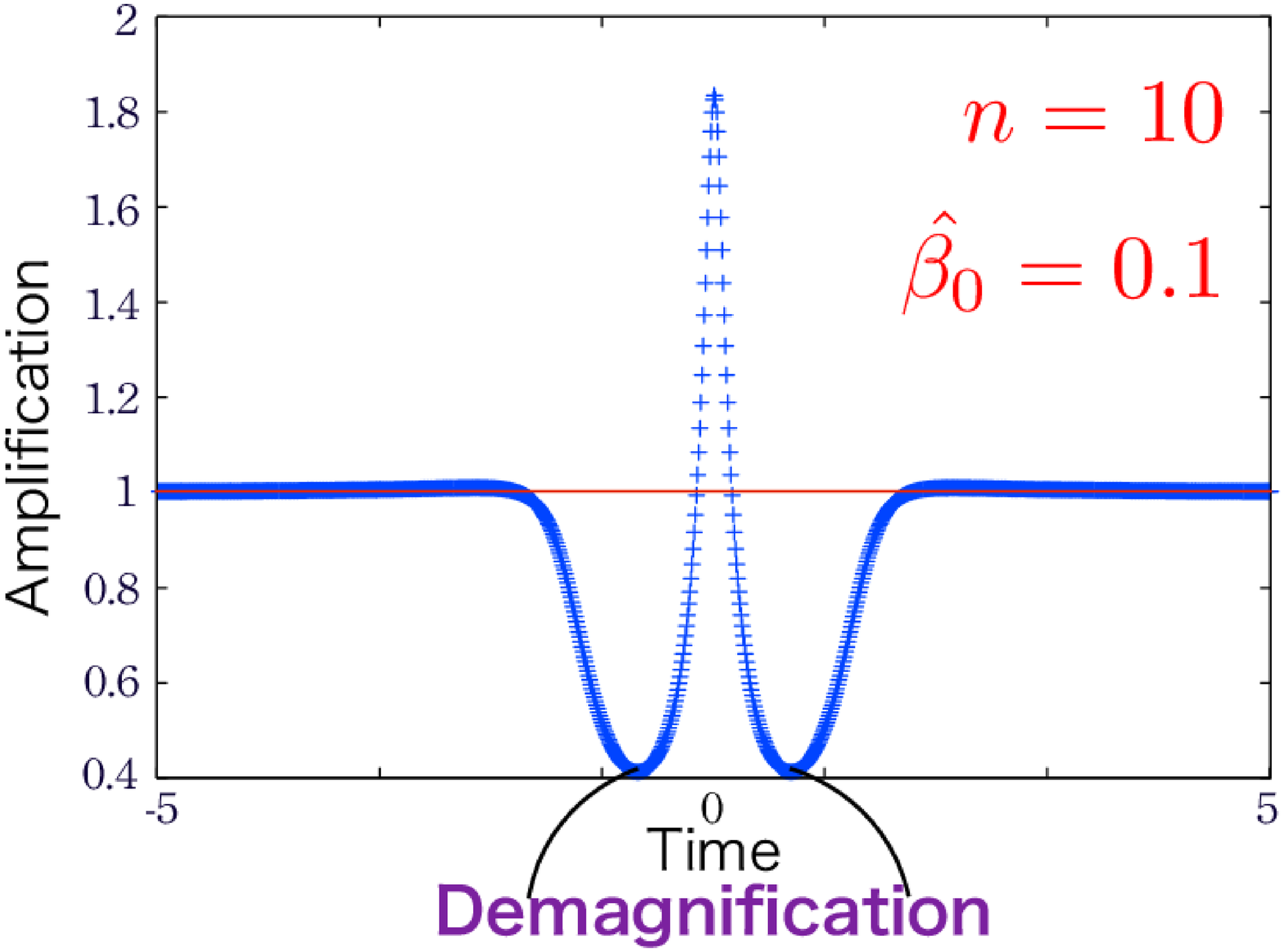}
\caption{
Numerical light curves for the same minimum impact parameter 
of the light trajectory $\hat\beta_0 = 0.1$. 
The source star moves at constant speed and 
the source position changes as 
$\hat\beta(t) = (\hat\beta_0^2 + t^2)^{1/2}$, 
where time is normalized by the Einstein ring radius crossing time. 
Top left, top right, bottom left, and bottom right panels 
correspond to $n=1, 2, 3$, and $10$, respectively. 
For convenience, a thin (red) line denotes $A_{\text{tot}}=1$. 
}
\label{figure-2}
\end{figure}

Figure \ref{figure-2} shows numerical light curves for $n = 1, 2, 3$, and $10$. 
As the power $n$ is larger, 
time-symmetric demagnification parts in the light curves 
become longer in time and larger in depth. 
Cases of $n=3$ and $10$ show 
maximally 
$\sim 10$ and $\sim 60$ percent depletion of the light, 
when the source position is $\hat\beta \sim 1.1$ and 
$\hat\beta \sim 0.7$, respectively. 

Before closing this section, we briefly mention an effective mass. 
A simple application of the standard lens theory \cite{SEF} 
suggests that the deflection ($\alpha = \bar\varepsilon/b^n$) 
and magnification studied here 
correspond to a convergence (scaled surface-mass density) 
of the form 
\begin{equation}
\kappa(b) = \frac{\bar\varepsilon (1-n)}{2} \frac{1}{b^{n+1}} . 
\label{kappa}
\end{equation}
For $n>1$, therefore, the effective surface-mass density 
of the lens object is interpreted as negative in the framework 
of the standard lens theory. 
This means that the matter (and energy) need to be exotic 
if $n>1$. 

\section{Discussion and Conclusion}
We examined a gravitational lens model inspired 
by modified gravity theories and exotic matter and energy. 
By using an asymptotically flat, static, and spherically symmetric 
spacetime model of which the metric depends on 
the inverse distance to the power of positive $n$, 
it was shown in the weak-field and thin-lens approximations that 
demagnifying gravitational lenses could appear, 
provided the impact parameter of light $\hat\beta$ 
and the power $n$ satisfy $\hat\beta > 2/(n+1)$ 
in units of the Einstein ring radius 
under a large-$n$ approximation. 

Therefore, time-symmetric demagnification parts in numerical light curves 
by gravitational microlensing may be evidence of an Ellis wormhole, but they do not always prove it. 
Such a gravitational demagnification of light 
might be used for hunting a clue of exotic matter and energy 
that are described by an equation of state more general than 
the Ellis wormhole case. 
Examples of $n=3$ and $10$ show 
maximally 
$\sim 10$ and $\sim 60$ percent depletion 
of the light, 
when the source position is $\hat\beta \sim 1.1$ and 
$\hat\beta \sim 0.7$, respectively. 
It is left to future work to perform a numerical campaign 
for the vast parameter space. 

The above gravitational demagnification of light occurs 
presumably because modified lenses could act as 
an effectively negative (quasilocal) mass on a particular light ray 
(through Ricci focusing). 
Regarding this issue, a more rigorous formulation is needed. 
It would be interesting to study a relation 
between the model parameter $n$ and vital modified gravity theories 
(or matter models with an exotic equation of state) 
and also to make an interpretation of the parameter $n$ 
in the framework of the theory of general relativity.

The analytical approximate solution in this paper is obtained 
at the linear order of $1/\hat\beta$ to discuss the total magnification. 
Tsukamoto and Harada \cite{Tsukamoto2} 
have studied the next order of $1/\hat\beta$
to discuss the signed magnification sums, namely  
the difference between the amplifications of two images.

We would like to thank F. Abe, T. Harada, S. Hayward, K. Nakao, 
and M. Visser for useful conversations. 
We wish to thank N. Tsukamoto for stimulating comments 
on the lecture (T.K.) on the present subject 
at ``WormHole Workshop 2012'' at Rikkyo University.

\end{document}